\documentstyle[aps,prb,preprint,tighten]{revtex}
\def\bk{{\bf k}}

\def\br{{\bf r}}
\def\bR{{\bf R}}
\def\om{\omega}
\def\bkom{(\bk,\omega)}

\def\em{{\cal E}_m}
\def\en{{\cal E}_n}

\def\fullint{\int_{-\infty}^{+\infty}d\omega}
\def\akw{A(\bk,\omega)}
\def\nk{n(\bk)}

\def\ybco{YBa$_2$Cu$_3$0$_7$}
\def\y_124{YBa$_2$Cu$_4$0$_8$}
\def\etal{{\it et al.}}
\begin{document}
\draft
%
%
\title{High $T_c$ Superconductors: \\
 New Insights from Angle-Resolved Photoemission}
\author{Mohit Randeria$^{1}$ and Juan-Carlos Campuzano$^{2,3}$}

\address{(1) Theoretical Physics Group, Tata Institute of Fundamental Research,
             Mumbai 400005, India \\
         (2) Physics Department, University of Illinois at Chicago,
             Chicago, IL 60607, USA \\
         (3) Material Science Division, Argonne National Laboratory,
             Argonne, IL 60439, USA 
         }

\address{%
\begin{minipage}[t]{6.0in}
\begin{abstract}
Recent angle-resolved photoemission (ARPES) studies of the high $T_c$ 
superconductors are reviewed. Amongst the topics discussed are: the 
spectral function interpretation of ARPES data and sum rules; studies 
of the momentum distribution and the Fermi surface (FS); dispersion of 
electronic states, flat bands and superlattice effects; unusual lineshapes 
and their temperature dependence; the question of bilayer splitting; 
detailed studies of the superconducting gap and its anisotropy; and, 
finally, studies of the pseudogap and evolution of the FS with doping 
in the underdoped materials. [Varenna Lectures, 1997].
\typeout{polish abstract}
\end{abstract}
\pacs{Key words: photoemission, high temperature superconductors}
\end{minipage}
}
\maketitle
\narrowtext

\noindent{\bf 1. Introduction} 

Angle-resolved photoemission spectroscopy (ARPES)
[\onlinecite{HUFNER}], [\onlinecite{REVIEWS}]
is a spectroscopy in which photons
are absorbed by the material and electrons are ejected out.
The control parameters are the frequency and polarization of the 
incident photons and the measured quantities are the kinetic
energy and the angle of emergence ($\theta,\varphi$) of the outgoing
electron relative to the sample normal. 

ARPES is now recognized as one of the major sources of 
insight into various aspects of the high temperature superconductors (HTSC).
This is remarkable, considering that
until ten years ago photoemission had never been used as a probe of
superconductivity. The first observation of the superconducting gap
in angle-integrated PES was given in [\onlinecite{IMER}], and
in angle-resolved form in [\onlinecite{OLSON_89}]. This was followed by
the observation of a normal state with a Luttinger Fermi surface 
[\onlinecite{CAMPUZANO_90}], [\onlinecite{OLSON_90}], 
flat bands in the dispersion [\onlinecite{GOFRON}],
[\onlinecite{DESSAU_93}], and the anisotropy of the SC gap
[\onlinecite{SHEN_93}]; see also [\onlinecite{SHEN_REVIEW}].
In these lectures, we will concentrate on the tremendous
progress of the past three years in which, we believe, there
has been a qualitative change in thinking about ARPES data 
and analyzing it. Along with this have come
a variety of new physics results which have shed very important
new light on the high Tc superconductors.

There are several reasons for the great success of 
ARPES for the high $T_c$ materials.  First, the great improvement 
in the experimental (especially energy) resolution 
allows one to study spectral features on
the scale of the SC gap in these materials. The FWHM of 
the energy resolution is routinely about 20 meV or larger.
The large gap energy scales and the higher $T_c$'s help considerably.
Second, Bi2212 has a natural cleavage plane in-between the BiO
bilayer which is believed to be van der Waals coupled leading
to the longest bond length in all the cuprates.  This results in
extremely smooth surfaces, with minimal charge transfer,
which are crucial for ARPES, since this is a
surface-sensitive technique due to the short escape depth 
($\sim 10 \AA$) of the outgoing electron.  
The third and final reason is the quasi-two-dimensionality of the 
electronic structure of the cuprates, which permits one to 
unambiguously determine the initial state momentum
from a final state measurement, since the 
component of $\bk$ parallel to the surface is conserved as the
electron emerges from the sample.

Nevertheless, there are important issues which have to be addressed.
We have little prior experience in analyzing ARPES data
on the energy scale of few 10's of meV. Much of the rest of these lectures
will focus on recent efforts towards properly analysing such data,
and developing an understanding of low frequency information contained
in it. We will focus here on the conceptual issues leaving many
of the technical details to the papers referred to in the text.

\medskip
\noindent{\bf 2. What does ARPES measure?}

The theoretical interpretation of
ARPES spectra is complicated by the fact that, in general,
photoemission measures a {\it nonlinear} 
response function. The photo-electron current at the detector
is proportional to the number of incident photons, i.e., 
to the {\it square} of the vector potential, and
the relevant correlation function is a three current correlation,
as first emphasized by Shaich and Ashcroft [\onlinecite{ASHCROFT}]. 

It is instructive to briefly review their argument. 
As in standard response function calculations,
lets look at an expansion of the current at the detector, the response, 
in powers of the applied vector potential (incident photons).
Let $\bR$ be the location of the detector in vacuum, and $\br$ 
denote points inside the sample. 
The zeroth order piece $\langle 0|j_\alpha (\bR,t)|0 \rangle$ vanishes
as usual; there are no currents flowing anywhere in the 
absence of the applied field. 
Here $|0\rangle$ is the ground state of the unperturbed system. 
The linear response also vanishes.
$\langle 0|j_\alpha (\bR,t)j(\br,t')_\beta |0 \rangle = 0$ and 
$\langle 0|j_\alpha (\br,t')j_\beta (\bR,t)|0 \rangle = 0$,
since there are no particles at the detector, 
in absence of the e.m. field, and $j_\beta (\bR,t)|0 \rangle = 0$.
Thus the leading term which survives is
\begin{eqnarray}
\nonumber
\langle j_\gamma (\bR ,t) \rangle \propto \int d\br' dt' d\br'' dt'' 
A_\alpha (\br',t')A_\beta (\br'' ,t'') \\
\langle 0|j_\alpha (\br',t')j_\gamma (\bR ,t)j_\beta (\br'', t'')|0 \rangle
\end{eqnarray}
where only current operators
{\it inside} the sample act on the unperturbed ground state on either side 
and the current at the detector is sandwiched in between.

The three current correlation function can be represented by the 
triangle diagrams [\onlinecite{CAROLI}] of Fig.~{1}, 
where the line between the two external photon vertices
is the Greens function of the ``initial state'' or ``photo-hole'' and the  
two lines connecting the photon vertex to the current at the detector
represent the ``photo-electron'' which is emitted from the solid.
There is a large literature [\onlinecite{PENDRY}]
on the ab-initio evaluation of the bare triangle diagram (A), 
incorporating the realistic electronic structure
and surface termination, together with multiple-scattering effects in
the photo-electron final state. Such studies are useful for understanding
the photoemission intensities, but not the line-shape and the many-body
aspects of the problem, which are the objects of primary interest for us.

All possible renormalizations -- vertex corrections and
self energy effects -- of the the bare triangle diagram are shown in
(B) through (F) of Fig.~{1}. It is easy to draw these
diagrams, but impossible to evaluate them in any controlled
calculation! Nevertheless, they are useful 
in understanding, qualitatively, what the various processes are
and in estimating their importance.
Diagram (B) represents the many-body renormalization of the 
occupied initial state that we are interested in; (C) and (D)
represent final state line-width broadening and inelastic scattering;
(E) is a vertex correction that describes the interaction of
the escaping photo-electron with the photo-hole in the solid,
and (F) is a vertex correction which combines features of
(D) and (E). [An additional issue in a quantitative theory of
photoemission is related to the modification of the external
vector potential inside the medium, i.e., renormalizations
of the photon line.]

Let us first discuss the validity of the sudden approximation,
for 15 - 30 eV (ultraviolet) incident photons,
by making some simple time scale estimates.
The question is: is the outgoing photo-electron sufficiently fast that 
one can safely ignore its interaction with the photo-hole?
The time $t$ spent by the escaping photo-electron in vicinity of photo-hole
is the time available for interactions (``vertex corrections'')
which would invalidate the sudden approximation. 
A photoelectron with a kinetic energy of (say) 20 eV
has a velocity $v = 3 \times 10^8$ cm/s. The relevant length scale, which
is the smaller of the screening radius (of the photo-hole) 
and the escape depth, is $\sim 10 \AA$.  
Thus $t = 3 \times 10^{-16}$ s. This is to be compared with the
time scale for electron-electron interactions (which are the
dominant source of interactions at the high frequencies
of interest): $t_{ee} \sim 2 \pi / \omega_p = 4 \times 10^{-15}$ s,
using a plasma frequency $\omega_p \simeq 1$ eV for the cuprates
(this would be even lower if c-axis plasmons are involved).
{\it If} $t \ll t_{ee}$, then we can ignore vertex corrections. 
Our very crude estimate is $t/t_{ee} \simeq 0.1$; all that we can say is
that the situation with regard to the validity of the impulse approximation
is not hopeless, but clearly a better estimate or a different approach
is needed (see below!).

Very similar estimates can be made for renormalizations
of the outgoing photo-electron due to its interaction
with the medium; again e-e interactions dominate at the
energies of interest. The relevant length scale here is the
escape depth, which leads to a process of self-selection:
those electrons that actually make it to the detector with
an appreciable KE have suffered no collisions in the medium.
Such estimates indicate that the ``inelastic background'' 
must be small -- although its precise dependence on
$\bk$ and $\om$ is uncertain.

To summarize: (1) the corrections to the sudden approximation are
probably small and we shall test its validity further below. 
(2) Final state line-width effects are negligible. A clear experimental
proof for Bi2212 is the fact that deep in the SC state a  
resolution limited spectral peak is seen, as discussed below in Section 6.
(3) While the additive extrinsic background due to inelastic
scattering is small, its precise form remains an important 
unresolved problem.

\medskip
\noindent{\bf 3. Spectral Functions and Sum Rules}

Assuming the sudden approximation and ignoring the extrinsic background,
the ARPES intensity, or the energy distribution curve (EDC), is given by
\begin{equation}\label{intensity}
I\bkom = I_0(\bk) f(\omega) A\bkom
\end{equation}
where $\bk$, the in-plane momemntum, gives the location in the 
2D Brillouin zone, and $\om$ is the energy of the initial state
measured relative to the chemical potential. (Experimentally
$\om$ is measured relative to the Fermi level of a good metal
like Pt or Au in electrical contact with the sample).
$I_0(\bk)$ includes all the  kinematical factors and 
the dipole matrix element (squared). It depends, in addition to
$\bk$, on the incident photon energy and polarization. The only 
general constraints on $I_0$ come from dipole selection rules
which we will discuss later on.

The spectral line-shape ($\om$ dependence) of the 
EDC and its $T$ dependence, at the low
frequencies and temperatures of interest to us, are entirely
controlled by $f(\omega)\akw$.
Here $\akw$  is the initial state or ``photo-hole'' spectral function
$\akw = (-1 / \pi){\rm Im}G(\bk,\om+i0^+)$, and the
the Fermi function $f(\om) = 1/[\exp(\om/T) + 1]$ ensures that
we are only looking at the {\it occupied} part of this
spectral function. This can formally be seen as follows:
the spectral function consists of two pieces 
$\akw = A_{+}\bkom + A_{-}\bkom$, which are spectral weights to 
add and to remove an electron from the system. In ARPES, where
one extracts an electron one is measuring
$A_{-}\bkom = \sum_{m,n} \left[ e^{-\em}/{\cal Z} \right]
|\langle n |c_\bk | m \rangle |^2 
\delta(\om + \en - \em)$, which can be rewritten,
using standard manipulations, as $A_{-}\bkom = f(\omega)\akw$. 

To gain confidence in the validity of a simple spectral function 
interpretation of ARPES data in the layered cuprates we have
used the following strategy [\onlinecite{RANDERIA}]. 
Let us assume that the sudden approximation is valid, deduce
some general consequences based on sum rules, and then test those
experimentally.  
The well-known sum rule $\int_{-\infty}^{+\infty} d\om \akw = 1$
is not very useful for ARPES since it is a sum-rule for
PES ($A_{-}$) {\it and inverse} PES ($A_{+}$).
The density of states (DOS) sum rule $\sum_\bk \akw = N(\om)$ 
is also not directly useful since there is the $\bk$-dependent matrix
element factor $I_0(\bk)$. Very recently, we have looked at the
photoemission DOS defined by $N_p(\om) = \sum_\bk I_0(\bk)\akw$
in attempt to simulate angle-integrated data by $\bk$-summing ARPES
data, with very interesting results and strong parallels with
STM data; the reader is referred to ref.~[\onlinecite{INPROGRESS}] for
this new development.

The important sum rule is
\begin{equation}\label{nofk}
\fullint f(\omega) \akw = n(\bk),
\end{equation}
which directly relates the energy-integrated ARPES intensity 
to the momentum distribution $\nk$. Somewhat surprisingly its 
usefulness has never been exploited in the ARPES literature. 
We will use (\ref{nofk}) to derive an approximate
sum rule valid at $\bk_F$ which we will use to test the validity
of the spectral function interpretation and, thereby, (indirectly)
of the sudden approximation. Later, having checked this 
we will use (\ref{nofk}) to get experimentally information on $n(\bk)$.

We first focus on the Fermi surface $\bk = \bk_F$. 
One of the major issues that will occupy us in the rest of these 
lectures, is how to define $\bk_F$, at finite temperatures, in a strongly
interacting system which may not even have well-defined 
quasiparticles -- and how to
to determine it experimentally. At this point, we simply define 
the Fermi surface to be the $\bk$-space locus 
of gapless excitations in the normal state,
so that $A(\bk_F,\om)$ has a peak at $\om = 0$.

To make further progress with eqn.~(\ref{nofk}), we need to
make a weak particle-hole symmetry assumption:
$A(\bk_F,-\omega) = A(\bk_F,\omega)$ for ``small'' $\omega$,
where ``small'' means those frequencies for which
there is significant $T$-dependence in the spectral function.
It then follows that [\onlinecite{RANDERIA}]
$\partial n(\bk_F)/\partial T = 0$, i.e.,
{\it the integrated area under the EDC at} $\bk_F$ 
{\it is independent of temperature}. 
To see this, rewrite eqn.~(\ref{nofk}) as
$n(\bk_F) = 1/2 - \int_0^\infty d\omega\tanh\left(\omega/2T\right)
\left[A(\bk_F,\omega) - A(\bk_F,-\omega) \right]/2$, and
take its $T$-derivative. It should be emphasized that
we cannot say anything about the {\it value} of $n(\bk_F)$,
only that it is $T$-independent. (A much stronger assumption, 
$A(\bk_F,-\omega) = A(\bk_F,\omega)$ for {\it all} $\omega$,
is sufficient to give $n(\bk_F)=1/2$ independent of $T$).
We emphasize the approximate nature of the $\bk_F$-sum-rule
since there is no exact symmetry that enforces it.

We note that we did not make any use of any properties of
the spectral function other than the weak p-h symmetry assumption,
and to the extent that this is also valid in the SC state, our
conclusion $\partial n(\bk_F)/\partial T = 0$ holds equally
well below $T_c$. There is the subtle issue of the meaning of
``$\bk_F$'' in the SC state. In analogy with the FS as the
``locus of gapless excitations'' above $T_c$, we can define
the ``minimum gap locus''  below $T_c$. We will describe this 
in great detail in Section 10 below; it suffices to note here that
``$\bk_F$'' is independent of temperature, within experimental
errors, in both the normal and SC state of the systems studied thus far. 

\medskip
\noindent{\bf 4. Experimental Details}

We now describe the experiments that first test the above ideas and 
then use them to get new information.
Most of the data to be discussed 
in these lectures is on very high quality 
single crystals of Bi$_2$Sr$_2$CaCu$_2$O$_{8+x}$ (Bi2212), grown by the 
traveling solvent floating zone method with an infrared 
mirror furnace, with low defect densities and sharp x-ray diffraction
rocking curves with structural coherence lengths $\sim 1250\AA$.
The near optimally-doped samples (which we shall focus on, except
in the last part of the lectures) have a $T_c = 87$K with
a transition width of 1K as determined by a SQUID magnetometer.
The samples are cleaved in-situ at 13 K in a vacuum of
$<$ 5x$10^{-11}$ Torr, and have optically flat surfaces
as measured by specular laser reflections.  
Another measure of the sample quality, within ARPES,
is the observation of ``umklapp" bands in the
electronic structure (described below) due to the presence of a structural
superlattice distortion. 

The experiments were performed at the Synchrotron Radiation Center,
Wisconsin, using a high resolution 4-m normal incidence monochromator with
a resolving power of $10^4$ at $10^{11}$ photons/s.
The samples are carefully oriented in the sample holder to an 
accuracy 1$^\circ$ by Laue diffraction, and the orientation is
further confirmed by the observed symmetry of sharp PES features
around high symmetry points. Various experiments have been done
using 17 -- 22 eV photons, with an energy resolution (FWHM) in the
range of 15 -- 25 meV and a typical momentum window of angular range 
$\pm 1^\circ$. 

For the Brillouin zone of Bi2212, we use a square lattice notation 
with the $\Gamma {\bar M}$ along the CuO bond direction. 
$\Gamma = (0,0)$, ${\bar M} = (\pi,0)$, $X=(\pi,-\pi)$ and $Y=(\pi,\pi)$
in units of $1/a^*$, where $a^* = 3.83\AA$ is the separation 
between near neighbor Cu ions. 
(The orthorhombic a axis is along $X$ and b axis along $Y$).

\medskip
\noindent{\bf 5. Experiments on Sum Rule and} $n(\bk)$

Fig.~{2}(a) shows ARPES spectra for a near-optimal Bi2212 
$T_c=87$ K at the FS crossing along $(\pi,0)$ to $(\pi,\pi)$ at 
two temperatures: $T=13$ K, which is well below $T_c$,  
and $T=95$ K, which is in the normal state. The two data sets
were normalized in the positive energy region [\onlinecite{SOL}],
which after normalization was chosen to be the common zero baseline.
For details, see ref.~[\onlinecite{RANDERIA}].
Remarkably, even though the spectra themselves
are very strongly $T$-dependent, their integrated
intensity in Fig.~{2}(b) is constant within experimental error bars 
(arising from the normalization), as predicted by the 
$\partial n(\bk_F)/\partial T = 0$ sum rule.
The sum rule has also been checked at other FS crossings, but is
much less informative along  the $(0,0)$ to $(\pi,\pi)$ crossing
where the observed line-shape is not too strongly $T$-dependent.

An important application of (\ref{nofk}) would
be to use it to experimentally determine the momentum distribution,
particularly since no other methods have successfully addressed this
problem for the cuprates (e.g., positron annihilation in \ybco\ 
apparently only yields information about the chain bands).
There are several caveats to keep in mind here, before discussing
the data. First, we do not have an absolute scale for the integrated
intensity, and, the unknown scale factor is $\bk$-dependent, since
from (\ref{intensity}) and (\ref{nofk}): 
$\int d\om I\bkom = I_0(\bk) n(\bk)$.
(In principle, electronic structure theory can provide useful input on
the $\bk$-dependence of matrix elements [\onlinecite{PENDRY}]).
Second, we do not know what the ``zero'' for the integrated intensity
is, in view of the unknown ``extrinsic background''. 
Finally there is the question of the integration limits in (\ref{nofk}):
while the Fermi function cutoff makes the upper limit irrelevant,
the lower limit may be more problematical. 
Thus, quantitative studies on $n(\bk)$ are not possible at the moment, 
but important qualitative information can be extracted as shown below. 

In view of the above discussion, we first choose to
illustrate the idea of measuring $n(\bk)$ on \y_124 
(Y124) [\onlinecite{GOFRON}], rather than on Bi2212. 
The spectra in Fig.~{3} (a) and (b) show intense peaks at low
energy which get cut off at high binding energies, for the occupied $\bk$
states, and also a loss of emission intensity once $\bk_F$ is crossed.
Thus the integrated intensity is not seriously affected either by the
background problem or uncertainties about the lower limit of integration.
The only drawback is the absence of SC within ARPES,
presumably due to surface problems, and the Y124 data (bulk $T_c = 82$K) 
are in a non-superconducting state at 12K. In Fig.~{3} (c) and (d),
we plot the integrated intensity:
the FS crossings along the S-Y-S direction deduced from the dispersion 
data in (a) are are indicated on the plot in (c). 
As discussed above, we get information about the momentum
distribution to the extent that we assume that the rapid $\bk$-variation 
comes from $n(\bk)$ near the FS, while the prefactor $I_0(\bk)$ is
slowly varying. 

The situation in Bi2212 (see Fig.~{4}) 
is not as clear cut, both as regards the
background, since there is considerable emission after crossing 
$\bk_F$, even though its much smaller than in the occupied states, and
as regards the lower limit of integration. 
Even with these limitations, the integrated intensity shown in 
Fig.~{4} is very informative [\onlinecite{CAMPUZANO_96}].
(Note that the integrated intensity for $\bk$ way past $\bk_F$, i.e.
deep on the unoccupied side, is set to zero, by hand).
To minimize the effects of the matrix elements and the slowly varying 
additive background, it is useful to look at peaks in 
$\vert\nabla_\bk n(\bk)\vert$. As seen from Fig.~{4}, these 
correlate very well with the FS crossing inferred from
the dispersion data.

\medskip
\noindent{\bf 6. ARPES Spectra: Qualitative Features}

At this stage, having gained some confidence in
interpreting ARPES spectra in terms of the (occupied part of the)
one-particle spectral function, let us discuss some of the
important qualitative features of the data. 
The first thing to emphasize is that the peak of the experimental
spectrum , the EDC, is {\it not} necessarily that of the 
spectral function. This is obvious from eqn.~(\ref{intensity}),
$I \sim f(\om)\akw$, and directly seen from the data in Fig.~{5}. 
In the normal state, the EDC peak is produced by the Fermi function 
$f(\om)$ cutting off the spectral function $\akw$, 
which would presumably peak at $\om = 0$ for $\bk = \bk_F$. 
We note, in passing, that recently we have succeeded
in developing methods for ``dividing out the Fermi function'' in the
data, which is nontrivial because of the convolution with the energy 
resolution. This gives very useful direct information about
$\akw$, as we will discuss elsewhere [\onlinecite{INPROGRESS}].

An important consequence is that the normal state spectral
function is extremely broad, the observed full width of the EDC
being less than the actual half-width of $\akw$!
Does this anomalous normal state spectrum imply a 
breakdown of Fermi liquid theory, as suggested by
numerous transport experiments?
In principle, one should be able to answer this question 
by analyzing the ARPES data using:
\begin{equation}
\akw = {{\Sigma''\bkom / \pi}\over
{\left(\om - \epsilon_\bk - \Sigma'\bkom\right)^2 + \Sigma''\bkom^2}}
\end{equation}
where $\Sigma'$ and $\Sigma''$ are the real and imaginary parts of
the self energy. In practice, the number of parameters involved in
$\Sigma$, coupled with uncertainties about the extrinsic background, 
lead to serious questions about the uniqueness of such fits.
Instead of trying to extract the $\om$-dependence of $\Sigma$
it is more useful, at the present time, to focus on the 
$\bk$-dependence of the line-shape as one approaches the FS.
For a Fermi liquid, with well defined quasiparticles, the spectrum 
should sharpen up as $\bk_F$ is approached. However, as the normal
state data in Fig.~{6} clearly show, 
and simple fits [\onlinecite{NFL_FITS}] corroborate, this does not happen. 
{\it There are no well defined quasiparticles above $T_c$}!
It is very important to emphasize that the large linewidths
observed in ARPES are not extrinsic, or artifacts of any analysis.
As we will see next, when quasiparticles do exist (for $T \ll T_c$)
they are clearly seen in the experiment.

The remarkable $T$-dependent changes in the line-shape
in Figs.~{2} and {5} may be understood as follows. 
For $T < T_c$ the SC gap opens up and spectral weight at $\bk_F$
shifts from $\om =0$ (in the normal state) to either side of it,
of which only the occupied side ($\om < 0$) is probed by ARPES.
At the lowest temperature the EDC peak {\it is} the peak of the
spectral function (unlike the normal state) since, as is obvious
from Fig.~{5}, the fermi function has now become sharper and
spectral weight has moved down to below the gap energy.
A detailed analysis of the SC gap data will described in 
Section 10 below.

Another striking feature of the data is the sharpening of the peak 
with decreasing $T$ in the SC state. This is {\it not} 
a ``BCS pile up'' in the density of states, a description
frequently used in the early literature, since we are not measuring a DOS. 
With a rapid decrease in linewidth below $T_c$,
the only way the conserved area sum rule can be satisfied is by having
a large rise in intensity.
The dramatic decrease in the linewidth ($\Sigma''$) below $T_c$ 
is a consequence of the the SC gap leading to a suppression of
electron-electron scattering which was responsible
for the large linewidth above $T_c$. {\it Thus coherent quasiparticle
(q.p.) excitations do exist for} $T \ll T_c$ [\onlinecite{K_DEPENDENCE}].
The rapid $T$-dependence of the line width 
is in qualitative agreement with  the results of various
transport measurements [\onlinecite{LINEWIDTH}].
A quantitative extraction of the scattering rate from ARPES 
data is an important open problem.

It is worth emphasizing that every aspect of this data, from
the broad normal state spectrum to the highly non-trivial
SC state line shape, points to the importance of e-e interactions.
The strong $T$-dependence of the linewidth is very unusual, and would not
occur in conventional metallic SC's where the e-e interaction
contribution to the scattering rate is weak.
We will argue below that e-e interactions are also responsible
for the non-trivial dip and hump structure (see Fig.~{2})
present beyond the sharp q.p. peak in the SC state; see Section 9.
Finally, the fact that we see spectral shifts in the same data
all the way down to 100 meV $\sim$ 1000 K, 
for a temperature change of 100 K, also suggests that
e-e interactions are at work.

\medskip
\noindent{\bf 7. Normal State of Optimally Doped Bi2212}

We now briefly summarize the main results of a very detailed study 
[\onlinecite{DING_96}] of the electronic excitations in
the normal state ($T = 95$K) of near-optimal Bi2212 ($T_c = 87$K).  
We begin with a discussion of the dispersion of the electronic excitations
and the Fermi Surface. Two representative data sets are plotted in
Fig.~{6}: the left panel shows dispersing peaks along the
diagonal $(0,0)$ to $(\pi,\pi)$, while the right panel shows data 
along the zone boundary $(\pi,0)$ to $(\pi,-\pi)$.
Spectral peak positions as a function of $\bk$ are plotted in 
Fig.~{7}(b), and the corresponding Fermi surface (FS)
crossings in Fig.~{7}(a).

In addition to the symbols in Fig.~{7}, there are also
several curves, which we now describe; these curves make clear the 
significance of all of the observed features. 
The thick curve is a 6-parameter tight-binding fit [\onlinecite{NORMAN_95a}]
to the $Y$-quadrant data; this represents the main CuO$_2$ band.
The two thin curves are obtained by shifting
the main band fit by $\pm{\bf Q}$ respectively, where
${\bf Q} = (0.21\pi, 0.21\pi)$ is
the superlattice (SL) vector known from structural studies 
[\onlinecite{SUPERLATTICE}].
We also have a few data points lying on a dashed curve,
which is obtained by shifting the ${\bf k}$ of the main band 
by $(\pi,\pi)$; this ``shadow band'' will be discussed below.
The Fermi surfaces corresponding to the main band fit (thick line),
the SL umklapps (thin lines) and the shadow band (dashed) are
are plotted as curves in Fig.~{7}(a).
We note that the main FS is a large hole-like barrel centered
about the $(\pi,\pi)$ point whose enclosed area
corresponds to approximately 1.17 holes per planar Cu (i.e.,
a hole doping of 0.17).
One of the key questions is why only one CuO main band is found in
Bi2212 which is a bilayer material. We will discuss this in depth in
the Section 9. 

The next important point relates to
the ``shadow bands'' first observed in ARPES experiments [\onlinecite{AEBI}]
done in a rather different mode (roughly, those experiments measure
$\int_{\delta\om} d\om \akw$ over a small range $\delta\om$ near
$\om = 0$). The shadow bands were not seen earlier in the EDC mode
experiments probably because of their sensitive photon energy
dependence and the absence of a strong feature near $E_F$.
These ``shadow bands'' were predicted early on
to arise from short ranged antiferromagnetic correlations [\onlinecite{KAMPF}].
An alternative explanation, which needs to be tested further, is that
they are of structural origin: Bi2212 has a face-centered orthorhombic
cell with two inequivalent Cu sites per plane, which by itself
could generate a $(\pi,\pi)$ umklapp.

We now turn to the effect of the superlattice (SL) on the ARPES
spectra.  This is very important, since a lack of understanding of
these effects led to incorrect conclusions regarding such basic
issues as one versus two Femi surfaces (see Section 9), and 
the anisotropy of the SC gap (see Section 10).
All of the experimental evidence is in favour
of interpreting the SL umklapp bands as arising from
a final state effect in which the exiting photo-electron scatters
off the structural SL superlattice distortion (which lives
primarily) on the Bi-O layer [\onlinecite{BIO_SL}].

We use the polarization selection rules 
to disentangle the main and SL bands in the $X$-quadrant
where the main and umklapp FSs are very close together; see Fig.~{7}(a).
The point is that $\Gamma X$ (together with the $z$-axis) and, similarly
$\Gamma Y$, are mirror planes, and an initial state
arising from an orbital which has $d_{x^2-y^2}$ symmetry about a
planar Cu-site is odd under reflection in these mirror planes.
With the detector placed in the mirror plane the final state
is even, and one expects a dipole-allowed transition when the photon
polarization ${\bf A}$ is perpendicular to (odd about)
the mirror plane, but no emission when the polarization is parallel to
(even about) the mirror plane.
While this selection rule is obeyed along $\Gamma Y$ it is violated
along $\Gamma X$. In fact this apparent violation of selection rules
in the X quadrant, was a puzzling feature of all previous
studies [\onlinecite{SHEN_REVIEW}] of Bi2212.
It was first pointed out in ref.~[\onlinecite{NORMAN_95b}], and
then experimentally verified in ref.~[\onlinecite{DING_96}],
that this ``forbidden'' $\Gamma X \vert\vert$ emission
originates from the SL umklapps.
We will come back to the $\Gamma X \vert\vert$ emission in the
superconducting state below.

\noindent{\bf 8. Extended Saddle Point Singularity}

Some aspects of the normal state dispersion
plotted in Fig.~{7}(b) deserve special mention:
while the dispersion along the diagonal $(0,0)$ to $(\pi,\pi)$
is very rapid, that near the $(\pi,0)$ point is very flat.
In particular, along $(0,0)$ to $(\pi,0)$ there is an
intense spectral peak is the main band, which disperses
towards $E_F$  but stays just below it at a binding energy of
(approximately) $-30$ meV. This is often called the
``flat band'' or ``extended saddle point'', and appears to
exist in all cuprates, though at different binding energies
in different materials [\onlinecite{GOFRON}],
[\onlinecite{DESSAU_93}], [\onlinecite{SHEN_REVIEW}].

In our opinion this flat band is not a consequence of the
bare electronic structure but rather a many-body effect. The
argument for this is that a tight-binding description of
such a dispersion requires fine-tuning (of the ratio of the next-near 
neighbour hopping to the near-neighbour hopping) which would be
unnatural even in one material, let alone many. 

Another important issue is whether this flat band leads to a singular
density of states.  It is very important to recognize that,
while Fig.~{7} (b) {\it looks like} a conventional band structure,
the dispersing states whose ``centroids'' or ``peak positions'' are plotted
are extremely broad, with width comparable to binding energy,
and these simply cannot be thought of as quasiparticles.
This general point is true at all $\bk$'s, but specifically for
the flat band region it has the effect of spreading out the spectral weight
over such a broad range that any singularity in the DOS 
would be washed out.

\medskip
\noindent{\bf 9. Bilayer Splitting?}

On very general grounds, one expects that the two CuO$_2$ layers in
a unit cell of Bi2212 should hybridize to produce two electronic
states which are even and odd under reflection in a mirror plane
mid-way between the layers. Where are these two states? Why did we
find only one main ``band'' and only one FS in Fig.{7}?

We have carefully checked the absence of a FS crossing for 
the main band along $\Gamma\bar{M}$ by studying the 
integrated intensity and its derivative $|\nabla_{{\bf k}}n({\bf k})|$ 
and found no sharp feature in $n({\bf k})$. 
Further the FS crossing that we do see near $(\pi,0)$ along
$\Gamma\bar{M}$ in Fig.{7}(a) is clearly associated with a
SL umklapp band, as seen both from the dispersion data 
in Fig.{7} (b) and its polarization analysis.
This FS crossing is only seen in the $\Gamma\bar{M}\perp$ (odd)
geometry both in our data and in earlier work [\onlinecite{DESSAU_93}]
(where it was erroneously identified as part of a second FS closed
round around $\Gamma$). Emission from the main $d_{x^2 - y^2}$ band, 
which is even about $\Gamma\bar{M}$, is dipole forbidden, and one
only observes a weak SL signal crossing $E_F$.
This clearly demonstrates that the bilayer splitting
of the CuO$_2$ states does not lead to two experimentally
resolvable Fermi surfaces.

It should be emphasized that this, by itself, is not
in contradiction with electronic structure 
calculations [\onlinecite{BAND_THEORY}].
Whether or not the two Fermi surfaces are resolvable
depends sensitively on the exact doping levels and on the presence
of Bi-O pockets, which are neither treated accurately in the theory nor
observed in the ARPES data. However, there {\it is} a clear prediction
from band theory: at $\bar M = (\pi,0)$, where both states are occupied
the bilayer splitting is the largest, of order 0.25 eV. 

The normal state spectrum at $\bar M$ is so broad
that it may be hard to resolve two states.
However, for $T \ll T_c$, when a sharp quasiparticle peak
is seen, the bilayer splitting should be readily
observable. For this one needs to interpret the
non-trivial line shape at $\bar M$ shown in Fig.~{8}:
with a dip [\onlinecite{SHEN_REVIEW}], [\onlinecite{DESSAU_DIP}]
in between the q.p. peak and a broad bump at 100 meV.
Probably the simplest interpretation would be (I) where
the bump is the second band, which is resolved below $T_c$ 
once the first band becomes sharp. The other alternative (II) is
that non-trivial line shape is due to many-body effects in a single
spectral function $\akw$.  To choose between these two hypotheses,
we exploit the polarization dependence of the matrix elements.
In case (I) there are two independent matrix elements which,
in general, should vary differently with ${\bf A}$, and thus
the intensities of the two features should vary independently.
While for case (II), the intensities of the two features should
scale together. In ref.~[\onlinecite{DING_96}] we found, by varying the
z-component of ${\bf A}$, evidence supporting hypothesis (II):
the q.p. peak, dip and bump are all part of a single spectral
function for Bi2212. The same conclusion can be quite independently
reached from the dispersion data in the SC and normal states
shown in ref.~[\onlinecite{NORMAN_97}]. Additional experimental
evidence against a two band interpretation of the dip
structure comes from tunneling [\onlinecite{JOHNZ}]. 

There are two important questions arising from this conclusion.
First, what causes this non-trivial line shape? 
The answer is the non-trivial $\om$-dependence of the self energy:
at low $\om$, $\Sigma''$ is suppressed by the opening of the
gap which leads to the q.p. peak, but at $\om \gg \Delta$, 
$\Sigma$ must recover its normal state behavior. This effect
is qualitatively able to account for the dip-bump structure
[\onlinecite{THREE_DELTA}]. A more quantitative description of the
SC line-shape is lacking at the present time; for some recent
progress in this direction, see [\onlinecite{NORMAN_97}].

The second question to ask is: what conspires to keep the two states
degenerate? Anderson [\onlinecite{ANDERSON}] had predicted that 
many-body effects within a single layer would
destroy both the quasiparticles and the coherent bilayer splitting 
{\it in the normal state}. 
But why the splitting should not be visible in the SC state, where
q.p.'s do exist, is not so clear.  Finally, it should be mentioned that,
in contrast to the Bi2212 case, there is some evidence for
bilayer-split bands in YBCO [\onlinecite{CAMPUZANO_90}], 
[\onlinecite{LIU}], [\onlinecite{GOFRON}], although this problem
needs further investigation.

\medskip
\noindent{\bf 10. Superconducting Gap and its Anisotropy}

In this Section, we will first establish how the SC gap manifests itself
in ARPES spectra, and then directly map out its variation with $\bk$
along the FS.  Since ARPES is the only available technique 
for obtaining such information, it has a played an 
important role [\onlinecite{SHEN_93}], [\onlinecite{DING_GAP}]
in establishing the $d$-wave order parameter in the high Tc superconductors
[\onlinecite{OP_SYMMETRY}]. 

In Fig.~{9} we show SC state spectra for Bi2212 for a
sequence of $\bk$'s. In the normal state these $\bk$'s
go from the occupied (top) to unoccupied (bottom) states, 
through $\bk_F$, as shown in Fig.~{4}. 
However, in the SC state the spectral peaks do not disperse
through the chemical potential, rather they first approach $\om =0$
and then recede away from it, as can be clearly seen from
Fig.~{9} (b). In comparing the normal and SC state data
in Figs.~{4} and {9} (which have different energy scales!), 
it is important to bear in mind the 
discussion in Section 6 based on Fig.~{5} that in the normal state
the EDC peak is caused by the Fermi function cut-off while for a 
gapped spectrum, the EDC peak is that of the spectral function.

There are several important conclusions to be drawn from Fig.~{9}.
First the bending back of the spectral peak, for $\bk$ beyond $\bk_F$,
is direct evidence for particle-hole mixing in the SC state; for
details see ref.~[\onlinecite{CAMPUZANO_96}].
The energy of closest approach to $\om=0$ is related to the SC gap 
that has opened up at the FS, and a quantitative estimate of this gap 
will be described below. The location of closest approach to $\om=0$
(``minimum gap'') coincides, within experimental uncertainties, with 
the $\bk_F$ obtained from the normal state $n(\bk)$ analysis of 
Fig.~{4}. It is important for later purposes to note that 
the ``minimum gap locus'', determined in this way, gives information
about the underlying FS (which is, of course, gapped below $T_c$).

In Fig.~{10}, we show the $T= 13$K EDCs for the 87K $T_c$ sample for 
various points on the main band FS in the $Y$-quadrant.
Each spectrum shown corresponds to the minimum observable
gap along a set of ${\bf k}$ points normal to the FS,
obtained from a dense sampling of ${\bf k}$-space [\onlinecite{DENSE}].
We used 22 eV photons in a $\Gamma Y\perp$ polarization,
with a 17 meV (FWHM) energy resolution, and a $\bk$-window of radius 
0.045$\pi/a^*$.

The simplest gap estimate is
obtained from the mid-point shift of the leading edge of Bi2212
relative to Pt in electrical contact with the sample.  
This has no obvious quantitative validity,
since the Bi2212 EDC is a spectral function while
the polycrystalline Pt spectrum (dashed curve in Fig.~{10})
is a weighted density of states whose leading edge is an 
energy-resolution limited Fermi function.
We see that the shifts (open circles in Fig.~{11}) indicate
a highly anisotropic gap which vanishes in the nodal directions,
and these results are qualitatively similar to ones obtained
from the fits described below.

Next we turn to modeling [\onlinecite{DING_GAP}], [\onlinecite{DING_95}]
the SC state data in terms of spectral functions.
It is important to ask how can we model the non-trivial line shape
(with the dip-bump structure at high $\om$) in the absence of a 
detailed theory, and, second, how do we deal with the extrinsic background?
We argue as follows:
in the large gap region near $(\pi,0)$, we see a linewidth collapse
for frequencies smaller than $\sim 3\Delta$ upon cooling well below $T_c$.
Thus for estimating the SC gap at the low temperature,
it is sufficient to look at small frequencies, and to focus on
the coherent (resolution limited) piece of the spectral function.
(Note this argument fails at higher temperatures, e.g., just below
$T_c$).  We model this coherent piece by the BCS spectral function
$A({\bf k},\omega)=
u_{\bk}^2 \Gamma/\pi\left[(\omega-E_{\bf k})^2+\Gamma^2\right] +
v_{\bk}^2 \Gamma/\pi\left[(\omega+E_{\bf k})^2+\Gamma^2\right]$
where the coherence factors are
$v_{\bk}^2=1-u_{\bk}^2={1\over2}(1-{\epsilon}_{\bf k}/E_{\bf k})$
and $\Gamma$ is a phenomenological linewidth.
The normal state energy ${\epsilon}_{\bf k}$ is measured from  $E_F$
and the Bogoliubov quasiparticle energy is
$E_{\bf k}=\sqrt{\epsilon_{\bf k}^2+\vert\Delta({\bf k})\vert^2}$,
where $\Delta({\bf k})$ is the gap function.
Note that only the second term in $\akw$, with the $v_{\bk}^2$-coefficient,
makes a significant contribution to the ARPES spectra.

The effects of experimental resolution are taken into account via
\begin{equation}
{\tilde I}({\bf k},\omega)= I_0 \int_{\delta\bk} d\bk'
\int_{-\infty}^{+\infty} d\omega' R(\omega - \omega') 
f(\omega') A(\bk',\omega')
\end{equation}
where $R(\omega)$, the energy resolution, is a normalized Gausian 
and $\delta\bk$ is the $\bk$-window of the analyzer.
In so far as the fitting procedure is concerned,
all of the incoherent part of the spectral function
is lumped together with the experimental background into one
function which is added to the ${\tilde I}$ above. 
Since the gap is determined by fitting the
resolution-limited leading edge of the EDC,
its value is insensitive to this drastic simplification.
To check this, we have made an independent set of fits
where we do not use any background fitting function,
and only try to match the leading edges, not the full spectrum.
The two gap estimates are consistent within a meV.
Once the insensitivity of the gap to the assumed background is
established, there are only two free parameters
in the fit at each $\bk$: the overall intensity $I_0$ and the
gap $|\Delta|$; the dispersion ${\epsilon}_{\bf k}$ is known
from the normal state study, the small linewidth $\Gamma$ is
dominated by the resolution.

The other important question is the justification for using
a coherent spectral function to model the rather broad EDC along 
and near the diagonal direction. We have found that such a description
is self-consistent [\onlinecite{DING_GAP}], [\onlinecite{DING_95}] 
(though perhaps not unique),
with the entire width of the EDC accounted for by the
large dispersion (of about 60 meV within our ${\bf k}$-window) 
along the zone diagonal.

The gaps extracted from fits to the spectra of Fig.~{10}
are shown as filled symbols in Fig.~{11}.
For a detailed discussion of the the error bars 
and also of sample-to-sample variations 
we refer the reader to ref.~[\onlinecite{DING_GAP}].
The angular variation of the gap obtained from the fits
is in excellent agreement with $|\cos(k_x) - \cos(k_y)|$ form.
The ARPES experiment cannot
of course measure the phase of the order parameter, but this result
is strongly suggestive of $d_{x^2-y^2}$ pairing.
Such an order parameter arises naturally in theories with
strong correlations and/or antiferromagnetic spin fluctuations
[\onlinecite{DWAVE}]. 

For completeness, we add few lines clarifying the earlier
observation of two nodes in the $X$-quadrant [\onlinecite{DING_95}],
and the related non-zero gap along $\Gamma X$ in the $\Gamma X \vert\vert$
geometry [\onlinecite{DING_95}], [\onlinecite{KELLY}].
It was realized soon afterwards that these observations were
related to gaps on the superlattice bands [\onlinecite{NORMAN_95b}],
and not on the main band. To prove this experimentally, the
$X$-quadrant gap has been studied in the $\Gamma X \perp$ geometry
[\onlinecite{DING_GAP}] and found to be consistent 
with $Y$-quadrant $d_{x^2-y^2}$ result described above.

\medskip
\noindent{\bf 11. Pseudogap in the underdoped materials}

We finally turn to one of the most fascinating recent developments --
pseudogaps -- in high $T_c$ superconductors in which ARPES has again 
played a major role [\onlinecite{DING_NATURE}], 
[\onlinecite{STANFORD_PGAP}], [\onlinecite{DING_97}]. 
Our discussion here will be rather brief as pseudogaps
will be the main topic of a companion set of lectures by one of us (M.R.),
where the reader will also find more detailed references on
the comparison of ARPES with other probes of the pseudogap, and
of various theoretical approaches [\onlinecite{MR_VARENNA}].

Up to this point we have discussed optimally doped Bi2212.
We now contrast this with the remarkable behaviour of
the underdoped materials, where $T_c$ is suppressed by lowering
the carrier (hole) concentration. See Fig.~{12} for a schematic 
phase diagram [\onlinecite{DING_NATURE}]. Underdoping
was achieved by adjusting the oxygen partial pressure during annealing
the float-zone grown crystals. These crystals also have
structural coherence lengths of at least 1,250$\AA$ as seen from
x-ray diffraction, and optically flat surfaces upon cleaving,
similar to the near-optimally doped $T_c$ samples discussed above.
(Actually, those samples are now believed to be slightly overdopded,
optimal doping corresponding to $T_c = 92$K). We denote the underdoped
samples by their onset $T_c$: the 83K sample has a transition width of 2K
and the highly underdoped 15K and 10K have transition widths $> 5$K.
(Other groups have also studied samples where underdoping was
achieved by cation substitution [\onlinecite{STANFORD_PGAP}]).

The first point to note about the high temperature 
ARPES spectra of underdoped Bi2212 is that they
become progressively broader with underdoping.
While the excitations of the optimally doped material
were anomalously broad (non-Fermi liquid behaviour), there
was nevertheless an identifiable spectral peak in the normal
state. In contrast, the underdoped spectra above $T_c$ are so
broad that there is no identifiable peak at all. One might
question: how do we know that these featureless EDCs are
spectral functions? There are two reasons: first,
even above $T_c$ there is observable dispersion, and 
second, way below $T_c$ a coherent (almost resolution-limited)
quasi-particle peak emerges (in the 83K $T_c$ samples in which this regime
is accessible).

In fact the SC state spectra in the underdoped regime look
very similar to those at optimal doping, with the one
difference that the spectral weight in the coherent q.p.
peak diminishes rapidly with underdoping. It is not possible
at the present time to quantify this important observation.
The ``minimum gap locus'' in the SC state (see Section 10)
suggests a large underlying Fermi surface, satisfying the Luttinger 
count of count of $(1+x)$ holes per planar Cu, and coincides with
the high temperature FS, which is the locus of gapless excitations
[\onlinecite{DING_97}]. The SC gap is found to be highly anisotropic,
with a node along the diagonal, and its $\bk$ variation along the 
FS is consistent with that of the optimally doped sample; 
see Fig.~{13}(a).

The major difference with the optimally doped sample is evident
upon heating through $T_c$. While the gapless excitations along
the diagonal remain gapless, the large gap along the $(\pi,0)$ to
$(\pi,\pi)$ crossing does not close above $T_c$, as seen from 
Fig.~{13}(b). One has to go to a (crossover scale) $T^*$ which
is much higher than $T_c$ before this gap vanishes and a closed
contour of gapless excitations (the FS) is recovered.
Note that in Fig.~{13} we use the leading edge shift to estimate
the gap, since except for $T \ll T_c$ we do not know enough 
about the line-shape to make any quantitative fits (as explained in 
Section 10).

It is important to emphasize that our understanding of the
83K $T_c$ sample is the best amongst all the underdoped materials.
In this sample all three regimes -- the SC state below $T_c$, 
the pseudogap regime (83K$= T_c < T < T^*$=170K) and the gapless 
``normal'' regime above $T^*$  -- have been studied in detail. 
In contrast, the 10K and 15K samples have such low $T_c$'s
and such high $T^*$'s that only the pseudogap regime is experimentally 
accessible. Nevertheless, the results on the heavily underdoped samples
appear to be a natural continuation of the weakly underdoped materials
and the results (similar, perhaps slightly larger [\onlinecite{UD}], 
magnitude of gap, higher value of $T^*$)
on the low $T_c$ samples are in qualitative agreement with  those
obtained from other probes (see ref.~[\onlinecite{MR_VARENNA}]).
Perhaps the most controversial of the results on the heavily underdoped
samples is the inference about a large underlying FS from the ``minimum
gap locus'' in the pseudogap regime [\onlinecite{DING_97}] as opposed
to small hole pockets. While this is certainly a tricky issue,
and there may also be materials problems in the very low $T_c$ sample,
we did not find any evidence for either the closure of a hole pocket
(concave arc about the $\Gamma$ point) or for shadow bands which are
$(\pi,\pi)$-foldbacks of the observed state.

To summarize the ARPES results in the underdoped regime: a highly
anisotropic SC gap is found in the underdoped samples which is
essentially independent of the doping 
level both in its magnitude [\onlinecite{UD}]
and in its $\bk$-dependence. Thus in this respect the underdoped samples
are very similar to optimally doped Bi2212. The key differences in the SC
state are first, the value of $T_c$, and second, the spectral weight 
in the coherent q.p. peak at $T \ll T_c$, both of which drop rapidly
with underdoping. Above $T_c$ the ARPES spectra in the underdoped state
are qualitatively different from optimal doping. ARPES continues to show
a gap which evolves smoothly through $T_c$ and has essentially
the same anisotropy as the SC gap. This suppression of spectral
weight, called the pseudogap, persists all the way to a much higher 
scale $T^*$ at which a locus of gapless excitations (Fermi surface)
is recovered. 

\medskip
\noindent{\bf 12. Conclusions}

In conclusion, we hope that we have been able to convey to the readers
the exciting new physics that has come out of ARPES studies of
the high $T_c$ superconductors. What is really astonishing is the
range of issues on which ARPES has given new insights:
from non-Fermi liquid behaviour with a Fermi surface, 
to the symmetry of the order parameter,
to the development of a Fermi surface in a doped Mott-insulator
and the pseudo-gap phenomena in the underdoped cuprates.

\medskip
\noindent{\bf Acknowledgements}

We gratefully acknowledge the contribution of all our collaborators,
and would particularly like to thank Hong Ding and Mike Norman
for the countless discussions which have shaped our understanding of 
the issues involved.  JCC was supported by the U. S. Dept. of Energy,
Basic Energy Sciences, under contract W-31-109-ENG-38, the National
Science Foundation DMR 9624048, and
DMR 91-20000 through the Science and Technology Center for
Superconductivity.

\begin{figure}
\caption{Diagrams contributing to the three-current correlation
formulation of the ARPES intensity. The dark lines are renormalized
propagators and the shaded blocks are vertex corrections. For simplicity
the arrows are shown in only (A).
The physical processes represented by each diagram are
discussed in the text.}
\label{1}
\end{figure}

\begin{figure}
\caption{
(a) ARPES spectra for Bi2212 $T_c=87$ K at $\bk = \bk_F$ (FS crossing
along $(\pi,0)$ to $(\pi,\pi)$) at $T=13$ K and $T=95$ K.
(b) Integrated intensity vs.~temperature showing that the area
is conserved.
}
\label{2}
\end{figure}

\begin{figure}
\caption{
(a) and (b): EDCs for \y_124 at 12 K for various $\bk$'s. 
Spectra labeled 1 through 13 are along the SYS direction and spectra
14 though 18 are along $\Gamma$Y.
(c) and (d): Integrated intensity, proportional
to $n(\bk)$, for the EDCs in (a) and (b).
The $\bk$ points are indicated in the BZ, the hatched
area denotes occupied states. 
The arrows show Fermi surface crossings
inferred from the dispersion.
The $\bk$-resolution of $\pm 1$ degree
corresponds to $\delta k_x a \simeq \pm 0.17$ as shown.
}
\label{3}
\end{figure}

\begin{figure}
\caption{ (a): Normal state ($T=95$K) Bi2212 spectra for a set
of $\bk$ values (in units of $1/a^*$).
(b): Integrated intensity (black dots) from data in (a) giving information
about the momentum distribution; its
derivative is shown by a solid curve (arbitrary scale).
}
\label{4}
\end{figure}

\begin{figure}
\caption{SC ($T=13$K) and normal state ($T=95$K) Bi2212 spectra
(solid curves) and reference Pt spectra (dashed curves) at the same
temperatures. This shows how the fermi function cutoff produces
the normal state EDC peak, but the peak in the SC state is
an intrinsic feature of the spectral function.
}
\label{5}
\end{figure}

\begin{figure}
\caption{Normal state (T=95K) spectra for Bi2212 along two symmetry lines
at values of the momenta shown as open circles in the upper insets.
The photon polarization, {\bf A}, is horizontonal in each panel.
}
\label{6}
\end{figure}

\begin{figure}
\caption{Fermi surface (a) and dispersion (b) obtained from normal state
measurements.  The thick lines are obtained by a tight binding fit to the
dispersion data of the main band with the thin lines $(0.21\pi,0.21\pi)$
umklapps and the dashed lines $(\pi,\pi)$ umklapps of the main band.  Open
circles in (a) are the data.  In (b), filled circles are for odd initial states
(relative to the corresponding mirror plane), open circles for even initial
states, and triangles for data taken in a mixed geometry.  The
inset of (b) is a blowup of $\Gamma X$.
}
\label{7}
\end{figure}

\begin{figure}
\caption{Low temperature (T=13K) EDC's of Bi2212 at $\bar{M}$ for various
incident photon angles.  The solid (dashed) line is $18^\circ$ ($85^\circ$) from
the normal.  The inset shows the height of the sharp peak for data
normalized to the broad bump, at different incident angles.
}
\label{8}
\end{figure}

\begin{figure}
\caption{Superconducting state EDCs for Bi2212
for the set of $\bk$-values (1/a units) which are shown at the top.
(For corresponding normal state data, see Fig.~{4}).
(b) SC state peak positions (white dots) versus $\bk$ for data of part (a).
The $\bk_F$ marked is the same as that determined from the normal state
analysis of Fig.~{4}.
}
\label{9}
\end{figure}

\begin{figure}
\caption{Bi2212 spectra (solid lines) for a 87K $T_c$ sample at 13K
and Pt spectra (dashed lines)
versus binding energy (meV) along the Fermi surface in the $Y$ quadrant.
The photon polarization and BZ locations of the data points
are shown in inset to Fig~{11}.
}
\label{10}
\end{figure}

\begin{figure}
\caption{$Y$ quadrant gap in meV versus angle on the Fermi surface
(filled circles) from fits to the data of Fig.~{10}. Open symbols
show leading edge shift with respect to Pt reference.
The solid curve is a d-wave fits to the filled symbols.
}
\label{11}
\end{figure}

\begin{figure}
\caption{
Schematic phase diagram of Bi2212 as a function of hole doping.
The filled symbols are the measured $T_c$'s for the superconducting
phase transition from magnetic susceptibility.
The open symbols are the $T^{*}$ at which the (maximum) 
gap seen in ARPES closes; for the $T_c=$ 10K sample the symbol at 301K
is a lower bound on $T^{*}$.
}
\label{12}
\end{figure}

\begin{figure}
\caption{
Momentum and temperature dependence of the gap estimated from 
leading edge shift (see text).
a) ${\bf k}$-dependence of the gap along the ``minimum gap locus''
(see text) in the 87K $T_c$, 83K $T_c$ and 10K $T_c$ samples, measured at
14K. 
b) $T$-dependence of the (maximum) gap in a near-optimal 87K 
sample (circles), underdoped 83K (squares) and 10K (triangles) samples.
Note smooth evolution of gap from SC to normal state for 83K sample.
}
\label{13}
\end{figure}


\begin{references}
\bibitem{SOL}
The essentially $\omega$-independent emission at positive frequencies,
which is not cut-off by the Fermi function, is due to higher harmonics of
the incident photon beam (``second order light'').
\bibitem{K_DEPENDENCE}
Actually, this has been established only for $\bk$'s
around $(\pi,0)$. We cannot answer the question of whether
quasiparticles exist for $T \ll T_c$ along the $(0,0)$ to $(\pi,\pi)$
FS crossing because of the rapid dispersion within our $\bk$-window
(see Fig.~{7} and also Section 10).
\bibitem{BIO_SL} 
This would allow us to understand (1) why polarization
selection rules are obeyed for the $\Gamma\bar{M}$ mirror plane,
and (2) qualitatively, why the intensities for odd (only main-band
contributes) and even (only the SL contributes)
polarizations along $\Gamma X$ are comparable. This point was not
fully appreciated at the time [\onlinecite{DING_96}] was published
\bibitem{UD}
Recent data clearly shows that, inspite of some sample-to-sample
variation, there is a systematic trend for the maximum SC gap to
increase with increasing underdoping.
\end{references}
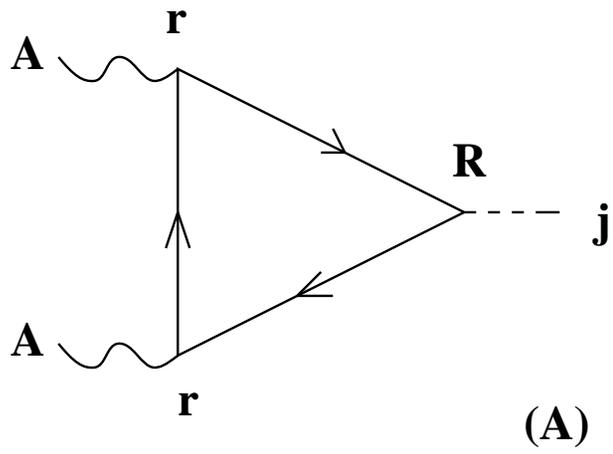
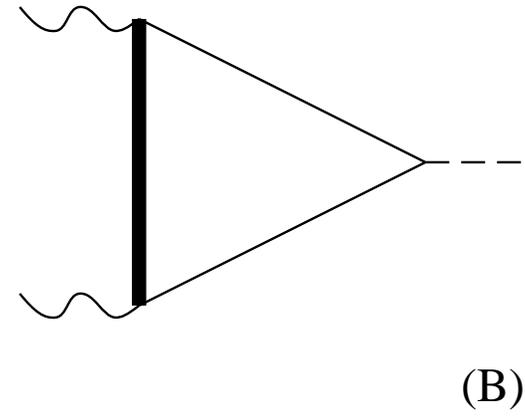
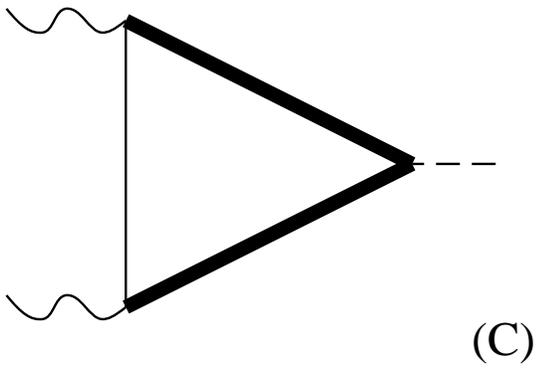
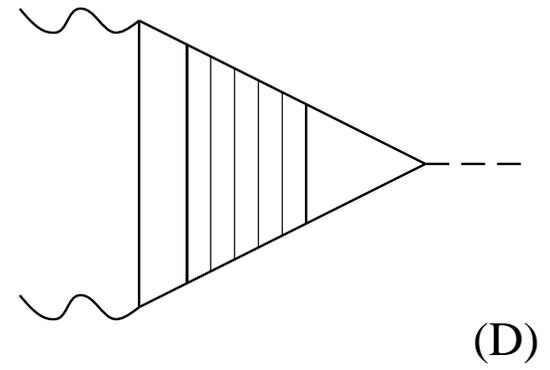
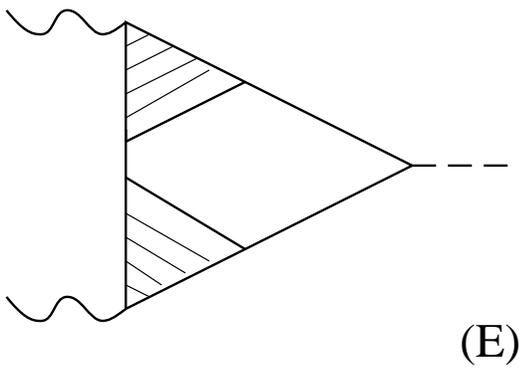
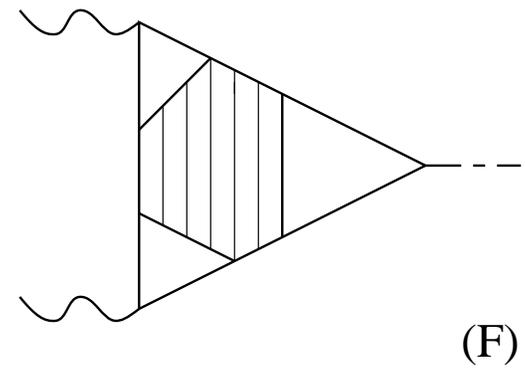
\end{document}